\documentclass[twocolumn,showpacs,floats,floatfix,superscriptaddress,aps,pra]{revtex4}
\usepackage{amsfonts}
\usepackage{amssymb}
\usepackage{amsmath}
\usepackage{calc}
\usepackage{graphicx}
\usepackage{bm}

\def\be{ \begin{equation} }
\def\ee{ \end{equation} }
\def\bea{ \begin{eqnarray} }
\def\eea{ \end{eqnarray} }
\def\bse{ \begin{subequations} }
\def\ese{ \end{subequations} }
\def\i{\red\,\text{i}\black}
\def\i{i}
\def\e{\,\text{e}}
\def\e{e}
\def\d{\,\text{d}}

\def\phase{\phi}
\def\U{\mathbf{U}}
\def\H{\mathbf{H}}

\def\c{\mathbf{c}}

\def\sech{\,\text{sech}}

\def\ket#1{\vert #1 \rangle}

\def\half{\tfrac12}

\def\pha{\phi}
\def\phb{\chi}
\def\phase{\alpha}

\def\tsum{\sum\nolimits}
\def\etal{\emph{et al. }}

\begin{document}

\author{Boyan T. Torosov}
\affiliation{Institute of Solid State Physics, Bulgarian Academy of Sciences, 72 Tsarigradsko chauss\'{e}e, 1784 Sofia, Bulgaria}
\author{Nikolay V. Vitanov}
\affiliation{Department of Physics, Sofia University, James Bourchier 5 blvd, 1164 Sofia, Bulgaria}
\title{High-fidelity error-resilient composite phase gates}
\date{\today }

\begin{abstract}
We present a method to construct high-fidelity quantum phase gates, which are insensitive to errors in various experimental parameters.
The phase gates consist of a pair of two sequential broadband composite pulses, with a phase difference $\pi+\phase/2$ between them, where $\phase$ is the desired gate phase.
By using composite pulses which compensate systematic errors in the pulse area, the frequency detuning, or both the area and the detuning, we thereby construct composite phase gates which compensate errors in the same parameters.
Particularly interesting are phase gates which use the recently discovered universal composite pulses, which compensate systematic errors in \emph{any} parameter of the driving field, which keep the evolution Hermitian (e.g., pulse amplitude and duration, pulse shape, frequency detuning, Stark shifts, residual frequency chirps, etc.).
%Because the composite phase gates can operate on resonance they are generally faster than dynamic, adiabatic and geometric phase gates.
%We study the performance of our gate in terms of fidelity and robustness.
%and find that it strongly outperforms the most commonly used dynamical phase gate,
\end{abstract}

\maketitle

%%%%%%%%%%%%%%%%%%%%%%%%%%%%%%%%%%%%%%%%%%%%%%%%%%%%%%%%%%%%%%%%%%%%%%%%%%%%%%%%%%%
%%%%%%%%%%%%%%%%%%%%%%%%%%%%%%%%%%%%%%%%%%%%%%%%%%%%%%%%%%%%%%%%%%%%%%%%%%%%%%%%%%%
%%%%%%%%%%%%%%%%%%%%%%%%%%%%%%%%%%%%%%%%%%%%%%%%%%%%%%%%%%%%%%%%%%%%%%%%%%%%%%%%%%%
\section{Introduction}

Quantum computers exploit the coherent superposition nature of quantum states and involve numerous phase-sensitive manipulations \cite{QI}.
%, single-qubit as well as multi-qubit ones \cite{QI}.
%For example, in quantum algorithms, such as Shor's factorization \cite{Shor} and Grover's search \cite{Grover}, one has to prepare such phase shifts very accurately. % \cite{Cirac}.
Insofar as quantum algorithms, such as Shor's factorization \cite{Shor} and Grover's search \cite{Grover}, involve a great number of phase gates, the accuracy of the latter is of crucial importance for high-fidelity quantum information processing.

As far as a phase shift of $\pi$ is concerned the simplest approach is to use a resonant $2\pi$ pulse that couples one of the qubit states with an ancilla state.
A variable phase shift $\phase$, however, requires a field with a suitable detuning, intensity and duration; such a variable phase shift is required, for example, for the construction of conditional quantum gates and the quantum Fourier transform \cite{QI}. %in deterministic quantum search \cite{Long01}.

There are two major types of phase gates: dynamic \cite{Cirac} and geometric \cite{geometric}.
The dynamic phase gate benefits from the simplicity of implementation because %, unlike the other phase gates,
 it requires just a single far-off-resonant pulsed field, which determines its wide-spread use.
The geometric phase gate has certain advantages in terms of robustness against parameter fluctuations, which come at the cost of more demanding implementations. % and imperfect fidelity.
An alternative phase gate uses adiabatic passage and relative laser phases \cite{laser phases}.

In the present paper we propose a different approach to construct a phase gate with an arbitrary phase.
Our method is based on the use of composite pulses (CPs) \cite{Levitt79,Freeman80,Levitt86,Freeman97}, which are a powerful method for quantum state control.
CPs combine high accuracy of manipulation with robustness to variations of the interaction parameters.
A CP is a sequence of pulses with suitably chosen relative phases.
These phases are used as control parameters to correct the errors which emerge in the interaction between a single pulse and a qubit.
The vast majority of CPs are designed to produce complete or partial population transfer in two-state or multi-state quantum systems.
Here we show how CPs can be used to produce well-defined phase shifts of the two states of a qubit.

The paper is organized as follows. 
First, we start with a brief overview of the theory of CPs and we show how CPs can be derived. 
Then we explain how CPs can be used to produce broadband (BB), adiabatic, and universal robust phase gates, which are insensitive to various types of experimental errors.
Then we test the performance of the proposed gates in terms of fidelity and robustness.

%%%%%%%%%%%%%%%%%%%%%%%%%%%%%%%%%%%%%%%%%%%%%%%%%%%%%%%%%%%%%%%%%%%%%%%%%%%%%%%%%%%
%%%%%%%%%%%%%%%%%%%%%%%%%%%%%%%%%%%%%%%%%%%%%%%%%%%%%%%%%%%%%%%%%%%%%%%%%%%%%%%%%%%
%%%%%%%%%%%%%%%%%%%%%%%%%%%%%%%%%%%%%%%%%%%%%%%%%%%%%%%%%%%%%%%%%%%%%%%%%%%%%%%%%%%
\section{Composite pulses}\label{CP}

Let us consider a two-state quantum system (a qubit), in a general state $\ket{\Psi}=c_1\ket{\psi_1}+c_2\ket{\psi_2}$, interacting with an external coherent field. Our goal is to create a phase gate, which is defined as an operation which changes the phase difference between $c_1$ and $c_2$ by some predefined amount $\phase$. In a matrix form it can be written as
\be\label{phase gate}
\Phi = \left[\begin{array}{cc}  e^{i\phase/2} & 0 \\  0 & e^{-i\phase/2} \end{array} \right].
\ee
In order to achieve this, we are going to use the powerful method of CPs.
To explain the idea of CPs, we first note that the evolution of our qubit is described by the Schr\"{o}dinger equation,
\be
i\hbar\partial_t\c(t)=\H(t)\c(t)\, ,
\ee
where $\c(t) = [c_1(t), c_2(t)]^T$ is a column vector with the probability amplitudes of the two states $|\psi_1\rangle$ and $|\psi_2\rangle$.
The Hamiltonian after the rotating wave approximation \cite{Bruce} reads
\be
\H(t) = (\hbar/2) \Omega(t) \e^{-\i D(t)} |\psi_1\rangle\langle \psi_2| + \text{h.c.},
\ee
with $D=\int_{0}^{t}\Delta(t^{\prime})\d t^{\prime}$, where $\Delta=\omega_0-\omega$ is the detuning between the field frequency $\omega$ and the Bohr transition frequency $\omega_0$.
The Rabi frequency $\Omega(t)$ is a measure of the field-system interaction:
 for laser-driven electric-dipole atomic transitions, $\Omega(t)=-\mathbf{d}\cdot\mathbf{E}(t)/\hbar$, where $\mathbf{E}(t)$ is the  laser  electric-field envelope and $\mathbf{d}$ is the transition dipole moment of the atom.
It is convenient to describe the evolution of the quantum system by means of the propagator $\U(t,t_i)$, which connects the probability amplitudes at any time $t$ to their initial values at time $t_i$: $\c(t)=\U(t,t_i)\c(t_i)$.
A general $2\times 2$ unitary propagator is parameterized by the Cayley-Klein parameters $a$ and $b$ as
\be\label{U}
\U = \left[\begin{array}{cc}  a  & b \\  -b^{\ast} & a^{\ast} \end{array} \right].
\ee
A constant phase shift $\pha$ in the driving field, $\Omega(t)\to\Omega(t)\e^{\i\pha}$, is mapped onto the propagator as
\be\label{Uph}
\U(\pha) = \left[ \begin{array}{cc} a & b \e^{\i\pha} \\ -b^{\ast}\e^{-\i\pha}  & a^{\ast} \end{array}\right].
\ee
If we have a sequence of $n$ identical pulses, each with a phase $\pha_k$, we obtain a CP whose effect upon the quantum system is described by the propagator
\be
\U^{(n)}=\U(\pha_n)\cdots\U(\pha_2)\U(\pha_1) .
\ee
If now the phases $\pha_k$ are chosen appropriately, the propagator $\U^{(n)}$ can be made much more robust to variations in the experimental parameters than the single-pulse propagator $\U$. This is the basic idea behind CPs and in such way one can produce a huge variety of broadband \cite{CP1,CP2}, narrowband \cite{NB}, and passband \cite{Kyoseva} CPs with respect to variations in essentially any experimental parameter.

%%%%%%%%%%%%%%%%%%%%%%%%%%%%%%%%%%%%%%%%%%%%%%%%%%%%%%%%%%%%%%%%%%%%%%%%%%%%%%%%%%%
%%%%%%%%%%%%%%%%%%%%%%%%%%%%%%%%%%%%%%%%%%%%%%%%%%%%%%%%%%%%%%%%%%%%%%%%%%%%%%%%%%%
%%%%%%%%%%%%%%%%%%%%%%%%%%%%%%%%%%%%%%%%%%%%%%%%%%%%%%%%%%%%%%%%%%%%%%%%%%%%%%%%%%%

\section{Composite phase gates: General}
%Let us cosider a two-state quantum system (a qubit) in a general state $\ket{\Psi}=c_1\ket{1}+c_2\ket{2}$, where $c_1$ and $c_2$ are the (complex-valued) probability amplitudes of the two-states $\ket{1}$ and $\ket{2}$.
%The phase gate, acting on such a qubit, is defined as an operation which changes the phase difference between $c_1$ and $c_2$ by some predefined amount $\phi$. In a matrix form it can be written as
%\be\label{phase gate}
%\Phi = \left[\begin{array}{cc}  e^{i\phi/2} & 0 \\  0 & e^{-i\phi/2} \end{array} \right]
%\ee
In this section we show how the phase gate \eqref{phase gate} can be constructed as a sequence of two CPs, with a fixed phase difference between them. 
In order to explain the idea, we first consider a simple sequence of two single pulses, wherein the second one has a relative phase $\chi$ with respect to the first one.
By using the single-pulse propagators \eqref{U} and \eqref{Uph} we find that for the sequence of two pulses, the total propagator is
\begin{align}\label{sequence}
\U_{\text{tot}}&=\U(\chi)\U(0)\notag\\
&=\left[\begin{array}{cc}  a^2-|b|^2 e^{i\chi} & a b + a^\ast b e^{i\chi}  \\  -a^\ast b^\ast -a b^\ast e^{-i\chi}& a^{\ast 2}-|b|^2 e^{-i\chi} \end{array} \right].
\end{align}
This simple expression reveals the idea of the phase gates proposed here:
 if each of the two single pulses is such that it causes complete population transfer ($a=0$ and hence $|b|=1$) and if we choose the phase $\chi$ of the second pulse to be equal to $\pi+\phase/2$, we obtain for the total propagator
$\U_{\text{tot}}=\U(\pi+\phase/2)\U(0)=\Phi$, which is exactly the phase gate defined by Eq.~\eqref{phase gate}.
This can be achieved by a sequence of two resonant pulses, for which the Cayley-Klein parameters are
$a=\cos A/2$, $b=-i\sin A/2$, where $A$ is the pulse area.
If we choose the area $A$ to be equal to $\pi$, we obtain $a=0$ and the total propagator is exactly the phase gate of Eq.~\eqref{phase gate}.
%\be
%\U_{\text{tot}}=\U_\pi(\pi+\phi/2)\U_\pi(0)=\Phi .
%\ee

This scheme, despite being quite simple and natural, has the same drawbacks as the single $\pi$ pulses, regarding the lack of robustness against variations of the interaction parameters. 
However, it is possible to overcome these shortcomings by replacing the two single $\pi$ pulses with two identical broadband CPs;
%In order to do this, we replace the two single pulses in the sequence \eqref{sequence} by two identical CPs;
 then the Cayley-Klein parameters $a$ and $b$ are determined by each of these CPs.
Each of the two CPs is robust against variations in one or more experimental parameters and hence, we obtain a robust phase gate.
Explicitly, the total propagator reads
\be\label{phase gate propagator}
\U_{\text{tot}}=\U_{\text{CP}_2}\U_{\text{CP}_1} ,
\ee
where
\bse\label{CP1&CP2}
\begin{align}
&\U_{\text{CP}_1}=\U(\pha_n)\cdots\U(\pha_2)\U(\pha_1), \\
&\U_{\text{CP}_2}=\U(\phb_n)\cdots\U(\phb_2)\U(\phb_1),
\end{align}
\ese
and
\be\label{phases second CP}
\phb_k = \pha_k+\pi+\phase/2.
\ee
Since the phases $\pha_k$ and $\phb_k$ are connected with this simple relation, from now on we will only refer to $\phi_k$.
The phases $\phi_k$ are chosen in such a way that they produce robust propagators $U_{\text{CP}_1}$ and $U_{\text{CP}_2}$, which are insensitive to variations in different experimental parameters.
It is straightforward to verify that if each of the two CPs of Eqs.~\eqref{phase gate propagator} and \eqref{CP1&CP2} produces complete population inversion up to the order $\epsilon^m$, i.e. $a=O(\epsilon^m)$, where $\epsilon$ is the deviation from the desired value of some interaction parameter,
 then the composite phase gate of Eq.~\eqref{sequence} will produce the desired target phase gate of Eq.~\eqref{phase gate} with the same accuracy $O(\epsilon^m)$.
In other words, the composite phase gate is accurate to the same order as the two composite $\pi$ pulses that it is composed of.

Depending on which type of CPs we use, we can produce different types of phase gates.
For example, a phase gate that is insensitive to errors in the pulse area can be produced by using BB CPs, which are robust against variations in the pulse area too.
In a similar way we can produce phase gates that are robust against any single parameter, or even such phase gates that are robust against \emph{all} experimental parameters.
% (e.g., Rabi frequency, pulse duration, static detuning, Stark shifts, frequency chirp, etc.).
In the next section, we will study these cases more closely.

In order to test the performance of our phase gates, we use the infidelity (failure) measure $F$, which we define as the Frobenius norm of the distance between the actual composite gate $\Phi^{\prime}$ and the desired phase gate $\Phi$ of Eq.~\eqref{phase gate},
\be
F = \sqrt{\tsum_{jk} \left|\Phi_{jk}^{\prime}-\Phi_{jk}\right|^2} .
\ee
%where $\Phi^{\prime}$ is the matrix of the composite phase gate and $\Phi$ is the target phase gate \eqref{phase gate}.

%%%%%%%%%%%%%%%%%%%%%%%%%%%%%%%%%%%%%%%%%%%%%%%%%%%%%%%%%%%%%%%%%%%%%%%%%%%%%%%%%%%
%%%%%%%%%%%%%%%%%%%%%%%%%%%%%%%%%%%%%%%%%%%%%%%%%%%%%%%%%%%%%%%%%%%%%%%%%%%%%%%%%%%
%%%%%%%%%%%%%%%%%%%%%%%%%%%%%%%%%%%%%%%%%%%%%%%%%%%%%%%%%%%%%%%%%%%%%%%%%%%%%%%%%%%

\section{Composite phase gates: Examples}

%%%%%%%%%%%%%%%%%%%%%%%%%%%%%%%%%%%%%%%%%%%%%%%%%%%%%%%%%%%%%%%%%%%%%%%%%%%%%%%%%%%
\subsection{Broadband composite phase gates}

%=================================================================
\begin{figure}[tb]
\includegraphics[width=8.5cm]{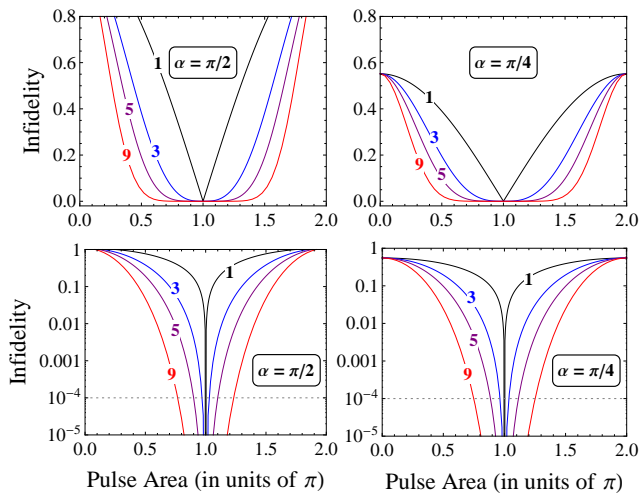}
\caption{Infidelity of the BB phase gate as a function of pulse area, for $n=1$, $n=3$, $n=5$ and $n=9$. The target phase is $\phase=\pi/2$ (left frames) and $\phase=\pi/4$ (right frames). The lower frames show the same infidelities as the upper frames, but in a logarithmic scale.}
\label{FidelityBB}
\end{figure}
%=================================================================

A phase gate, which is robust against errors in the pulse area, is produced by using a sequence of two area-compensating CPs.
Numerous such CPs have been proposed and demonstrated in the literature \cite{Levitt86}.
We use here the symmetric resonant CPs that we have derived recently \cite{CP1}; their phases are given by the analytic formula \cite{note}
\be\label{BBphases}
\pha_k=k(k-1)\frac{\pi}{n} \quad (k=1,2,\ldots,n) ,
\ee
where $n$ is the number of pulses, used in the CP.
Explicitly, the phases of the first few CP pulses are (modulo $2\pi$)
\bse
\begin{align}
& \left( 0,\tfrac23,0 \right)\pi, \\
& \left( 0,\tfrac25,\tfrac65,\tfrac25,0 \right)\pi, \\
& \left( 0,\tfrac27,\tfrac67,\tfrac{12}7,\tfrac67,\tfrac27,0 \right)\pi, \\
& \left( 0,\tfrac29,\tfrac23,\tfrac43,\tfrac29,\tfrac43,\tfrac23,\tfrac29,0 \right)\pi.
\end{align}
\ese
The phases of the second CP are found from here and Eq.~\eqref{phases second CP}.
These phases allow to suppress the error in the transition probability % produced by a $\pi$ pulse
 up to order $\mathcal{O}(\epsilon^{2n})$ where $\epsilon$ is the deviation of the pulse area $A$ from $\pi$, $A = \pi(1+\epsilon)$.
Obviously, the total number of pulses needed for the composite phase gate is $2n$.
%The relative phase between the two CPs is, as already explained, $\pi+\phase/2$.
As an example, for $n=3$, the total number of pulses is $2n=6$ and the relative phases are
%\bse
%\begin{align}
%&\pha_k = (0,\tfrac{2}{3}\pi,0) ,\\
%&\phb_k = (\pi+\half\phi,\tfrac{5}{3}\pi+\half\phi,\pi+\half\phi) ,
%\end{align}
%\ese
\be
\left(0,\tfrac{2}{3}\pi,0 ,\pi+\half\phase,\tfrac{5}{3}\pi+\half\phase,\pi+\half\phase\right) .
\ee
%where the phases $\phi_k$ are taken modulo $2\pi$.

Figure~\ref{FidelityBB} shows the infidelity of our composite phase gate as a function of the pulse area of the pulses in the sequence for target phases $\phase=\pi/2$ and $\phase=\pi/4$.
These phase gates are the most widely used ones in quantum information processing \cite{QI}. 
We see that by increasing the number of pulses in the two CPs, which compose the phase gates, the phase gates get increasingly robust to variations in the pulse area.

%%%%%%%%%%%%%%%%%%%%%%%%%%%%%%%%%%%%%%%%%%%%%%%%%%%%%%%%%%%%%%%%%%%%%%%%%%%%%%%%%%%
\subsection{Adiabatic composite phase gates}

%=================================================================
\begin{figure}[tb]
\includegraphics[width=7cm]{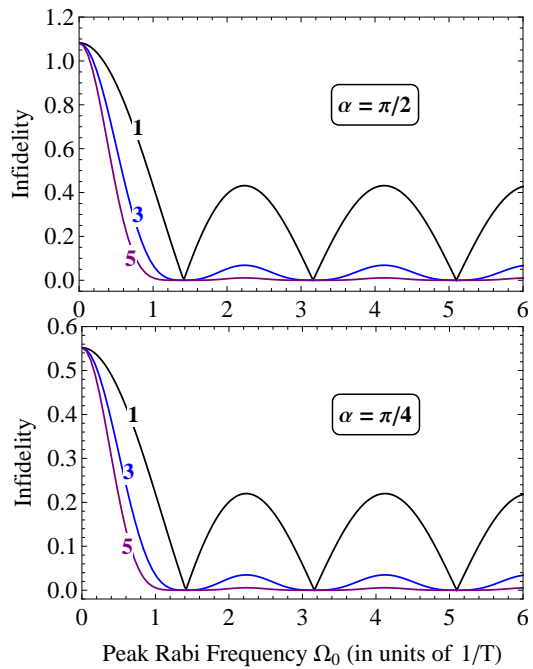}
\caption{Infidelity of the adiabatic phase gate for $n=3$ and $n=5$, compared with the infidelity, produced by a sequence of two single (non-composite) chirped pulses.
%The pulse shape is a hyperbolic secant with a peak Rabi frequency $\Omega_0$, the chirp is a hyperbolic tangent, with a rate of $1/T$, where $T$ is the pulse duration.
%
We use hyperbolic-secant pulses, $\Omega(t)=\Omega_0\sech(t/T)$, and hyperbolic-tangent detuning, $\Delta(t)=B\tanh(t/T)$, where $\Omega_0$ is the peak Rabi frequency, $B$ is the chirp rate and $T$ is the pulse duration. We have used chirp rate $B = 1/T$.
The target phase is $\phase=\pi/2$ (top frame) and $\phase=\pi/4$ (bottom frame).}
\label{FidelityCAP}
\end{figure}
%=================================================================

It was demonstrated in \cite{CP2} that CPs can be used to improve the adiabatic passage by using a sequence of phase-shifted chirped pulses.
It was shown that the composite phases do not depend on the particular shape of the pulse and the chirp, but only demand symmetric Rabi frequency and antisymmetric detuning.
The analytic formula for the phases is the same as the one for the BB pulses, Eq.~\eqref{BBphases}.
In Fig.~\ref{FidelityCAP} we compare the infidelities of the composite adiabatic phase gates for $n=3$ and $n=5$ with the single-pulse adiabatic phase gate ($n=1$).
We conclude that the composite adiabatic phase gate is extremely robust against variations in the peak Rabi frequency, which is due to the high fidelity and robustness of the composite adiabatic passage \cite{CP2}.

%%%%%%%%%%%%%%%%%%%%%%%%%%%%%%%%%%%%%%%%%%%%%%%%%%%%%%%%%%%%%%%%%%%%%%%%%%%%%%%%%%%
\subsection{Detuning-compensated composite phase gates}

%=================================================================
\begin{figure}[tb]
\includegraphics[width=7cm]{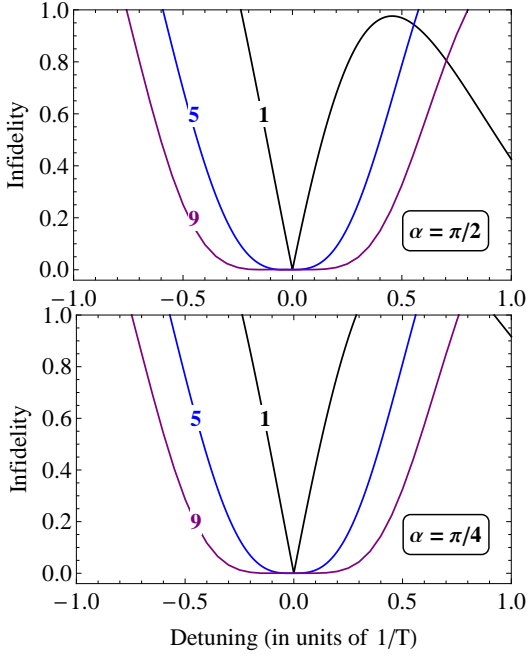}
\caption{Infidelity of the detuning-compensated phase gate for a sequence of two single hyperbolic secant pulses ($n=1$) and for a sequence of two CPs with $n=5$ and $n=9$. The composite phases are given by Eq.~\eqref{CP detuning}. 
The target phase is $\phase=\pi/2$ (top frame) and $\phase=\pi/4$ (bottom frame).}
\label{FidelityDelta}
\end{figure}
%=================================================================

Another major type of CPs are the ones which stabilize the excitation profile with respect to the frequency detuning from exact resonance.
These can be used to construct detuning-compensated composite phase gates.
We have derived earlier \cite{CP1} several detuning-compensated CPs,
\bse\label{CP detuning}
\begin{align}
&\left( 0, \tfrac13\pi, 0\right), \\
&\left(  0,0.747,0.424,0.747,0 \right)\pi, \\
&\left( 0,1.308,1.153,1.251,0.562,1.251,1.153,1.308,0 \right)\pi.
\end{align}
\ese
The first, second and third CPs of these produce a unit transition probability around single pulse areas of $\pi$, $3\pi/5$, and $4\pi/9$, respectively.

The detuning-compensated composite phase gates are obtained by using Eqs.~\eqref{phases second CP} and \eqref{CP detuning}.
For example, the six-pulse phase gate reads
\be
\left(0,\tfrac{1}{3}\pi,0 ,\pi+\half\phase,\tfrac{4}{3}\pi+\half\phase,\pi+\half\phase\right) ,
\ee

In Fig.~\ref{FidelityDelta} we plot the simulated infidelity of such gates.
It is evident from the figure that the phase gates are robust against variations in the detuning, which makes the method very useful for applications in situations, when exact resonance is not possible (for instance, in the presence of inhomogeneous broadening or Doppler shift).
%
%We note that, while we are using a sequence of $\pi$ pulses in the detuning-compensated CPs, this is not a restriction of our method and it is also possible to use pulses with smaller area.
%For instance, it was demonstrated in Ref.~\cite{CP1} how detuning compensation can be achieved with five $3\pi/5$ pulses or with nine $4\pi/9$ pulses, which allows the total area to be substantially reduced.

If we want to achieve compensation of errors in more than one parameter, we need to use CPs, which are robust against variations in several parameters, for instance pulse area and detuning \cite{CP1}. In the next subsection we examine the cases when a more general compensation of errors is possible.

%%%%%%%%%%%%%%%%%%%%%%%%%%%%%%%%%%%%%%%%%%%%%%%%%%%%%%%%%%%%%%%%%%%%%%%%%%%%%%%%%%%
\subsection{Universal composite phase gates}

\begin{table}[tb]
\caption{Universal CPs for complete population inversion, which compensate errors in arbitrary field parameters \cite{Universal}.}
\begin{tabular}{ll}\hline
CP & $(\pha_1,\pha_2,\ldots,\phi_n)$ \\ \hline
U3 & $(0,\frac12,0)\pi$ \\
U5a & $(0, \frac56, \frac13, \frac56, 0)\pi$\\
U5b & $(0, \frac{11}6, \frac13, \frac{11}6, 0)\pi$\\
U7a & $(0, \frac{11}{12}, \frac{5}{6}, \frac{17}{12},\frac{5}{6},\frac{11}{12},0)\pi$ \\
U7b & $(0, \frac{23}{12}, \frac{5}{6}, \frac{5}{12},\frac{5}{6},\frac{23}{12},0)\pi$ \\
%U9a & $(0, 0.635, 1.35, 0.553, 0.297,0.553,1.35,0.635,0)\pi$ \\
%U9b & $(0, 1.635, 1.35, 1.553, 0.297,1.553,1.35,1.635,0)\pi$ \\
U13a & $(0, \frac{3}{8}, \frac{42}{24}, \frac{11}{24}, \frac{8}{24}, \frac{37}{24}, \frac{2}{24},\frac{37}{24},\frac{8}{24},\frac{11}{24},\frac{42}{24},\frac{3}{8},0)\pi$ \\
U13b & $(0, \frac{33}{24}, \frac{42}{24}, \frac{35}{24}, \frac{8}{24}, \frac{13}{24}, \frac{2}{24},\frac{13}{24},\frac{8}{24},\frac{35}{24},\frac{42}{24},\frac{33}{24},0)\pi$ \\
\hline
\end{tabular}\label{table:universal}
\end{table}

%=================================================================
\begin{figure}[tb]
\includegraphics[width=8cm]{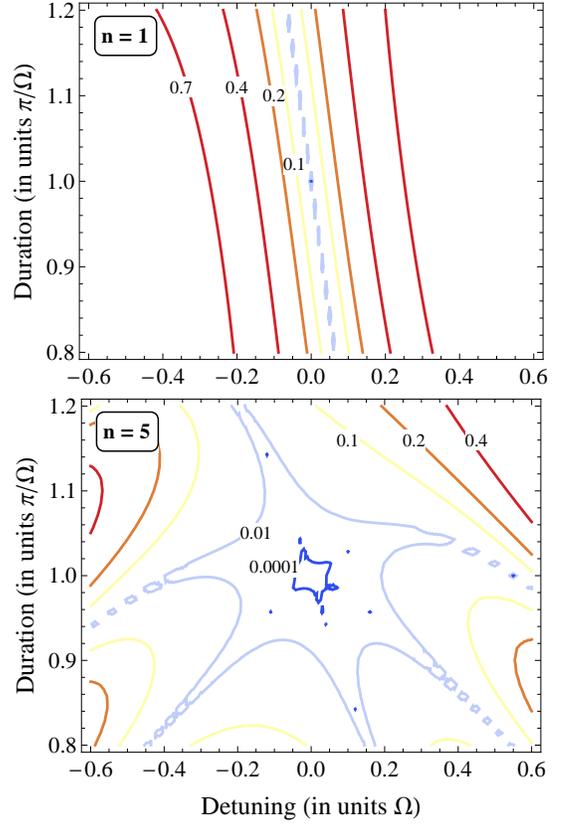}
\caption{Infidelity of the universal phase gate for (top frame) a sequence of two rectangular pulses and (bottom frame) a sequence of two CPs with $n=5$. The composite phases are $\pha_k=(0, 11, 2, 11, 0)\pi/6$. The target phase is $\phase=\pi/4$.}
\label{FidelityUNI}
\end{figure}
%=================================================================

Recently, Genov \etal \cite{Universal} derived CPs, which compensate errors in any parameter of the driving field.
This is done by using the most general parametrization \eqref{U} of the propagator, without any assumptions for the properties of the constituent pulses and their parameters, which justifies the term ``universal'' CPs.
The phases of the lowest-order universal CPs are given in Table \ref{table:universal}.
By using these universal CPs, we can build a universal phase gate, which is insensitive, up to a certain order, to any experimental error of the driving field.
%As an example, we can use the U5b universal CP \cite{Universal}, with phases
%\be
%\phi_k=(0, 11, 2, 11, 0)\pi/6 ,
%\ee
%to obtain a universal phase gate, insensitive to any kind of experimental errors \cite{note2}.
For example, the six- and ten-pulse composite phase gates read
\bse
\begin{align}
&\text{UPh6} = \left(0,\tfrac{1}{2}\pi,0 ,\pi+\half\phase,\tfrac{3}{2}\pi+\half\phase,\pi+\half\phase\right) ,\\
&\text{UPh10a} = \left(0, \tfrac56\pi, \tfrac13\pi, \tfrac56\pi, 0, \pi+\half\phase, \tfrac{11}6\pi+\half\phase, \right. \notag\\
&\qquad\qquad\qquad \left.\tfrac43\pi+\half\phase, \tfrac{11}6\pi+\half\phase, \pi+\half\phase\right) ,\\
&\text{UPh10b} = \left(0, \tfrac{11}6\pi, \tfrac13\pi, \tfrac{11}6\pi, 0, \pi+\half\phase, \tfrac{5}6\pi+\half\phase, \right. \notag\\
&\qquad\qquad\qquad \left.\tfrac43\pi+\half\phase, \tfrac{5}6\pi+\half\phase, \pi+\half\phase\right) ,
\end{align}
\ese
where \emph{a} and \emph{b} refer to different universal CP solutions \cite{Universal}. 

In Fig.~\ref{FidelityUNI}, we test the performance of the universal phase gates by plotting the infidelity as a function of the pulse duration and the detuning.
%In the upper frame, the phase gate is built by a sequence of two single rectangular pulses, and in the bottom frame, U5b CPs are used.
We see that by using the single-pulses approach, it is almost impossible to achieve high fidelity, while universal CPs deliver quite large high-fidelity areas of $F<0.01$, and even areas of ultrahigh fidelity $F<10^{-4}$.

%%%%%%%%%%%%%%%%%%%%%%%%%%%%%%%%%%%%%%%%%%%%%%%%%%%%%%%%%%%%%%%%%%%%%%%%%%%%%%%%%%%
%%%%%%%%%%%%%%%%%%%%%%%%%%%%%%%%%%%%%%%%%%%%%%%%%%%%%%%%%%%%%%%%%%%%%%%%%%%%%%%%%%%
%%%%%%%%%%%%%%%%%%%%%%%%%%%%%%%%%%%%%%%%%%%%%%%%%%%%%%%%%%%%%%%%%%%%%%%%%%%%%%%%%%%
\section{Conclusions}

We have presented an approach to construct high-fidelity error-resistant composite phase gates for quantum information processing.
These phase gates are formed by two identical composite $\pi$ pulses, the second of which is shifted by phase $\pi+\phase/2$ with respect to the first.
The properties of these CPs are directly transferred to the composite phase gate.
For example, a composite $\pi$ pulse, which compensates errors in the pulse area to a certain order $\epsilon^m$, produces a composite phase gate, which compensates errors in the pulse area to the same order $\epsilon^m$.
Of special interest are the universal phase gates, which compensate errors in all field parameters (pulse duration, pulse amplitude, detuning, unwanted chirp, Stark shift, etc.).
We note that these universal CPs have the huge practical advantage that they simultaneously compensate all kinds of errors, even unknown ones, as far as the evolution is coherent and hence unitary (no losses to other states).
However, if we wish robustness against a single parameter, e.g. pulse area or detuning, then better performance is provided by the dedicated CP, which compensates this particular error.

%Because the composite phase gates can operate on resonance they are generally faster than dynamic, adiabatic and geometric phase gates.

%%%%%%%%%%%%%%%%%%%%%%%%%%%%%%%%%%%%%%%%%%%%%%%%%%%%%%%%%%%%%%%%%%%%%%%%%%%%%%%%%%%%%%%%%%%%%%%%%%%%%%%%%%%%%%%%%%%%%%%%%%%%%%%%%%%
\acknowledgments

This work has been supported by the EC Seventh Framework Programme under grant agreement No. 270843 (iQIT), the Alexander-von-Humboldt Foundation and the Bulgarian National Science Fund grant DMU-03/103.

%%%%%%%%%%%%%%%%%%%%%%%%%%%%%%%%%%%%%%%%%%%%%%%%%%%%%%%%%%%%%%%%%%%%%%%%%%%%%%%%%%%%%%%%%%%%%%%%%
%%%%%%%%%%%%%%%%%%%%%%%%%%%%%%%%%%%%%%%%%%%%%%%%%%%%%%%%%%%%%%%%%%%%%%%%%%%%%%%%%%%%%%%%%%%%%%%%%
%%%%%%%%%%%%%%%%%%%%%%%%%%%%%%%%%%%%%%%%%%%%%%%%%%%%%%%%%%%%%%%%%%%%%%%%%%%%%%%%%%%%%%%%%%%%%%%%%
%%%%%%%%%%%%%%%%%%%%%%%%%%%%%%%%%%%%%%%%%%%%%%%%%%%%%%%%%%%%%%%%%%%%%%%%%%%%%%%%%%%%%%%%%%%%%%%%%
%%%%%%%%%%%%%%%%%%%%%%%%%%%%%%%%%%%%%%%%%%%%%%%%%%%%%%%%%%%%%%%%%%%%%%%%%%%%%%%%%%%%%%%%%%%%%%%%%


\begin{thebibliography}{99}

\bibitem{QI} M. A. Nielsen and I. L. Chuang, \emph{Quantum Computation and Quantum Information} (Cambridge University Press, 1990).

\bibitem{Shor}  P. W. Shor, in Proceedings of the 35th Annual Symposium on the Foundations of Computer Science, edited by
S. Goldwasser (IEEE Computer Society, Los Alamitos, 1994), p. 124;
P. W. Shor, SIAM J. Sci. Stat. Comput. \textbf{26}, 1484 (1997).

\bibitem{Grover} L. K. Grover, Phys. Rev. Lett. \textbf{79}, 325 (1997).

\bibitem{Cirac} T. Calarco, D. Jaksch, J. I. Cirac and P. Zoller, J. Opt. B \textbf{4}, 430 (2002).

\bibitem{geometric} M. V. Berry, Proc. R. Soc. London, Ser. A \textbf{392}, 45 (1984);
R. G. Unanyan, B. W. Shore, and K. Bergmann, Phys. Rev. A \textbf{59}, 2910 (1999);
R. Unanyan, M. Fleischhauer, B.W. Shore, K. Bergmann, Opt. Commun. \textbf{155}, 144 (1998);
A. Ekert, M. Ericsson, P. Hayden, H. Inamori, J. A. Jones, D. K. L. Oi and V. Vedral, J. Mod. Opt. \textbf{47}, 2501 (2000).

\bibitem{laser phases} X. Lacour, S. Gu\'erin, N. V. Vitanov, L. P. Yatsenko, and H. R. Jauslin, Opt. Commun. \textbf{264}, 362 (2006);
%Implementation of single-qubit quantum gates by adiabatic passage and static laser phases
H. Goto and K. Ichimura, Phys. Rev. A \textbf{70}, 012305 (2004).

\bibitem{Levitt79} M. H. Levitt and R. Freeman, J. Magn. Reson. \textbf{33}, 473 (1979).

\bibitem{Freeman80} R. Freeman, S. P. Kempsell, and M. H. Levitt, J. Magn. Reson. \textbf{38}, 453 (1980).

\bibitem{Levitt86} M. H. Levitt, Prog. Nucl. Magn. Reson. Spectrosc. \textbf{18}, 61 (1986).

\bibitem{Freeman97} R. Freeman, Spin Choreography (Spektrum, Oxford, 1997).

\bibitem{Bruce} B. W. Shore, \emph{The Theory of Coherent Atomic Excitation} (Wiley, New York, 1990).


\bibitem{CP1} B. T. Torosov and N. V. Vitanov, Phys. Rev. A \textbf{83}, 053420 (2011).

\bibitem{CP2} B. T. Torosov, S. Gu\'{e}rin, and N. V. Vitanov, Phys. Rev. Lett. \textbf{106}, 233001 (2011);
D. Schraft, T. Halfmann, G. T. Genov, and N. V. Vitanov, %Experimental demonstration of composite adiabatic passage
Phys. Rev. A \textbf{88}, 063406 (2013).

\bibitem{NB}
S. S. Ivanov and N. V. Vitanov, Opt. Lett. \textbf{36}, 1275 (2011);
%High-fidelity local addressing of trapped ions and atoms by composite sequences of laser pulses
N. V. Vitanov, Phys. Rev. A \textbf{84}, 065404 (2011).
%Arbitrarily accurate narrowband composite pulse sequences

\bibitem{Kyoseva}
E. Kyoseva and N. V. Vitanov, Phys. Rev. A \textbf{88}, 063410 (2013).
%Arbitrarily accurate passband composite pulses for dynamical suppression of amplitude noise

\bibitem{note} The formula presented here is simpler than the one in the cited paper, but leads to an identical result.

\bibitem{Universal} G. T. Genov, D. Schraft, T. Halfmann, and N. V. Vitanov,  arXiv:1403.1201v2 (2014).



%\bibitem{femto} J.-C. Diels and W. Rudolph, \textit{Ultrashort Laser Pulse Phenomena: Fundamentals, Techniques, and Applications on a Femtosecond Time Scale} (San Diego, Academic, 1996);
%M. Wollenhaupt, V. Engel and T. Baumert, Annu. Rev. Phys. Chem. \textbf{56}, 25 (2005).
%Brixner T, Pfeifer T, Gerber G, Wollenhaupt M and Baumert T 2005 Femtosecond Laser Spectroscopy ed P Hannaford (New York: Springer) chapter 9

%\bibitem{Long01} G.L. Long, Phys. Rev. A \textbf{64}, 022307 (2001).


%\bibitem{NB} N. V. Vitanov, Phys. Rev. A \textbf{84}, 065404 (2011).

%\bibitem{PB} E. Kyoseva and N. V. Vitanov Phys. Rev. A \textbf{88}, 063410 (2013).

%\bibitem{note2} For $n=5$, the phases of the universal CP U5b are the same as those of the detuning-compensated CP.



\end{thebibliography}
\end{document}